\documentclass[twocolumn]{aa}
\usepackage{graphicx}
\begin{document}
   \title{Optical variability of \object{PKS 0736+017}}

   \author{A. Ram\'\i rez \inst{1}, J.A. de Diego \inst{1}, D. Dultzin-Hacyan       \inst{1}
          \and
          J.N. Gonz\'alez--P\'erez \inst{2,}\inst{3}
          }

   \institute{Instituto de Astronom\'\i a, UNAM,
              Apdo. 70-264, Ciudad Universitaria, 04510 M\'exico, D.F.
              (M\'exico).\\
              \email{aramirez,jdo,deborah@astroscu.unam.mx}
         \and
             Universit\"at Hamburg, Hamburger Sternwarte, Gojenbergsweg 112, 21029, Hamburg, Germany
          \and Instituto de Astrof\'\i sica de Canarias, 38200 La Laguna, Tenerife, Spain\\
             \email{st1h317@hs.uni-hamburg.de}
             }

   \date{Received 6 October 2003 / Accepted 29 November 2003}

    \abstract{We present BVR photometric observations of the blazar \object{PKS 0736+017}. These observations were carried out with three telescopes in Mexico and two in Spain between December 1998 and April 2003. \object{PKS~0736+017} shows remarkable variation at different timescales and amplitudes. Maximum brightness was detected on December 19, 2001 (B=14.90$\pm$0.01, V=14.34$\pm$0.01, and R=13.79$\pm$0.01). A peculiar tendency to redden with increased brightness was detected throughout our observations. Moreover, in one season a good correlation between flux level and spectral slope is shown.
This {\it anomalous} behaviour cannot be described by common flare models of blazars. The flux vs. spectral slope correlation observed in this and other blazars is worth further study.

         }

\authorrunning{Ram\'\i rez et al.}

   \maketitle
%

\section{Introduction}

Blazars, as a class, exhibit flux variability at timescales from minutes to years (e.g. Ghosh et al. \cite{ghosh}; Vagnetti et al. \cite{vagnetti}; Garcia et al. \cite{garcia}; Brown et al. \cite{browna}, \cite{brownb}; Wright, McHardy \& Abraham \cite{wright98}), have high and variable polarization (Angel \& Stockman \cite{angel}), and core dominated radio structures with weaker extended radio emission (Antonucci \& Ulvestad \cite{antonucci}). The continuum emission  is characterized by a flat radio and steep optical-infrared spectrum, and the emission may extend to gamma-ray energies (Maraschi et al. \cite{maraschi}). Relativistic beaming of the jet emission in radio-loud active galactic nuclei is generally accepted as the explanation for the observed properties of blazars (Blandford \& Rees \cite{blanford}). Their strong highly variable polarization and fluctuating brightness are usually taken to reflect processes at the heart of the nuclear engine.

Some scenarios have been proposed to explain the rapid variability. Relativistic shock waves interacting with the inhomogeneous medium of the jet of blazars (Maraschi et al. \cite{maraschi89}; Qian et al. \cite{qian}; Marscher, Gear \& Travis \cite{marscher92}; Marscher \cite{marscher}; Gear et al. 1986; Ghisellini, Celotti \& Costamante \cite{ghisellini02}) have been invoked; additional models refer to flares generated on the accretion disk of a supermassive black hole or to eclipsing of hot spots by the accretion disk (Mangalam \& Wiita 1993; Wiita \cite{wiita}). The size of the emitting region is calculated using causality arguments.

The amplitudes of variability have been found to be correlated with the redshift, and anticorrelated with luminosities of AGNs (Hook et al. \cite{hook}; Cristiani et al. \cite{cristiani}; Di Clemente et al. \cite{diclemen}). With regard to timescales, no correlation was found with the amplitude of the variation in blazars by Nair (\cite{nair}) or with luminosity or redshift in AGNs (Cristiani et al. \cite{cristiani}). An anticorrelation between timescales and luminosity of radio-loud quasars was found by Netzer et al. (\cite{netzer}).

Many authors have found that the amplitude of the variations of blazars is higher at high frequencies, leading to a flattened spectrum when the flux increases, and to a steeper one when it decreases (Racine \cite{racine}; Gear, Robson \& Brown \cite{gear86}; Massaro et al. \cite{massaro}; Ghisellini et al. \cite{ghisellini}; Maesano, Massaro \& Nesci \cite{maesano}; Maesano, Montagni, Massaro \& Nesci \cite{maesano97}). However,  Ghosh et al. (\cite{ghosh}) suggest that it may not be correct to make such generalizations. They found occasionally that the amplitudes of the variations are similar in V and R, in the object \object{PKS 0235+164}, or even larger in R, in the object \object{PKS~0735+178}.

\object{PKS 0736+017} was classified as a blazar by Angel \& Stockman (\cite{angel}), and shows strong permitted and forbidden emission lines in the optical domain (Falomo, Scarpa \& Bersanelli \cite{falomo}). Padovani \& Urry (\cite{padovani}) and  Wall \& Peacock (\cite{wall}) reported \object{PKS 0736+017} as a Flat Spectrum Radio Quasar (FSRQ), while Brown et al. (\cite{browna}) report it as an Optically Violently Variable quasar, located at the centre of an elliptical galaxy, possibly a cD galaxy (Wright et al. \cite{wright98}; Wyckoff, Gehren \& Wehinger \cite{wyckoff}; Smith et al. \cite{smith}; Hutchings \cite{hutchings}; Falomo \& Ulrich \cite{falomo00}; Hughes et al. \cite{hughes}; Kotilainen, Falomo \& Scarpa \cite{kotil}), which has two close resolved faint companions that are embedded in the nebulosity of the object (Dunlop et al. \cite{dunlop}; Huthchings, Johnson \& Pyke \cite{hutchings88}; Bondi et al. \cite{bondi}). The absolute magnitude of the host galaxy has been determined between $M_R$=-24.3 and $M_R$=-22.2 (Falomo \& Ulrich \cite{falomo00}; Hutchings \cite{hutchings}), and $M_V$=-21.7 (Smith et al. \cite{smith}).

Historically, the brightness of this object has been variable at different  timescales. Pica et al. (\cite{pica}, \cite{pica88}) reported a 0.7 mag fade in six years and two flares of as much as 0.7 mag in 1974 and 1978, at timescales of weeks. Ghosh et al. (\cite{ghosh}) observed PKS 0736+017 in the R band.  They report variations of small amplitude ($\Delta$R=0.34$\pm$0.05) over timescales of six days. They also report a variation of 0.65$\pm$0.05 mag during the interval of its observations (between November 1995 and March 1998). Garcia et al. (\cite{garcia}) reported a variation of 0.8 mags (in V filter) in 400 days. McGimsey et al. (\cite{mcgimsey}) observed optical variation of 0.5 mag on a timescale of weeks and months, with an overall range of about 1.0~mag. Katajainen et al. (\cite{katajainen}) detected brightness variations in V between 16.25 and 16.84 magnitudes on timescales of 18 days. Schaefer (\cite{schaefer}) reported $\Delta m_B \simeq$0.5 mags in 1 day. Ram\'\i rez et al. (\cite{ramirez}) reported a variation of $\sim$ 0.3 mag in 4.5 hours in the optical band. Recently, Clements, Jenks \& Torres (\cite{clements}) found a flare with $\Delta$R=1.3 mags in two hours early in 2002.

Pica et al. (\cite{pica88}) detected that \object{PKS 0736+017} was at its minimum in 1981, but immediately increased, reaching its peak brightness in 1984. These authors reported an average magnitude of 16.8 at the faint and 15.45 at the bright stages, with an average magnitude of 16.05 (in $m_P$). The average magnitude for this object is between 16 and 17 mag in the V band (e.g. Katajainen et al. \cite{katajainen}; Garcia et al. \cite{garcia}; Wright et al. \cite{wright}; Bondi et al. \cite{bondi}).

Bondi et al. (\cite{bondi}) detected variability at radio frequencies  at two different  epochs (in early 1980 and late 1981). They found that the source was structured by two components and that the variability was caused by changes in both the flux densities and the separation of the components. In a third epoch, they identified the brightest component with the core. McAdam (\cite{mcadam}) reported flux variability at 408 MHz, and Kedziora-Chudczer et al. (\cite{kedziora}) observed polarization variations.

On the other hand, other studies were carried out without positive results.
Padrielli et al. (\cite{padrielli}) made multifrequency observations of PKS~0736+017 to search for correlation between low and high frequencies. Jones et al. (\cite{jones}) observed it as a part of their study of twenty radio sources to try to categorize flux variations due to changes in source scale, structure or the slope of the electron energy spectrum. Sitko, Schmidt \& Stein (\cite{sitko}) searched for optical polarimetry.

Brown et al. (\cite{brownb}), within their investigation of the variability properties from centimetre to ultraviolet wavelengths, reported that they did not find any correlation between near-infrared flux level and spectral slope for PKS~0736+017. However, they noted that excluding the data for the minimum, the flux shows a significant anticorrelation (more than 80\% confidence level) between flux level and spectral slope (taken $f_{\nu} \propto \nu ^{\alpha}$).

In this paper we present and discuss optical variability of the Optical Violently Variable (OVV) quasar \object{PKS~0736+017}; in \S 2 the observations and data reduction procedures are described; in \S 3 the variability behaviour is analyzed. Finally, in \S 4 we present our discussion and in \S 5 our conclusions.

\section{Observations and data reduction}

The observations were carried out with five telescopes in Mexico and Spain. Observations from Mexico were made with three telescopes: a 2.1 m (M1), a 1.5 m (M2), and a 0.84 m (M3) operated by the Observatorio Astron\'omico Nacional at San Pedro M\'artir, Baja California. Telescopes in Spain were the 1 m Jacobus Kapteyn Telescope (S1) operated by the Isaac Newton Group at the Observatorio del Roque de los Muchachos, and the 1.52 m (S2) Estaci\'on de Observaci\'on de Calar Alto (EOCA) at Centro Astron\'omico Hispano-Alem\'an Calar Alto (CAHA) operated by the Observatorio Astron\'omico Nacional. All the observations were made with BVR Johnson series filters. The log of the observations is given in Table 1. In this table the date and telescope used are listed in columns (1) and (2), respectively. Magnitudes and uncertainties for each filter are listed in columns (3)-(8).

The CCDs used with M1 and M2 were a Site SI003 with 1024X1024 pixels of $24\mu m \times 24 \mu m$ with a Metachrome II and VISAR coating to improve the response in the blue. With M3 the detector was a Thomson TH7398M CCD with 2048X2048 pixels, of $14\mu m \times 14 \mu m$, with a Metachrome II coating. S1 had attached a TEK1024AR: 1024X1024 pixels of $24\mu m \times 24 \mu m$, thinned AR coated, and S2 had a Tektronics TK1024AB: 1024X1024 pixels of $24\mu m \times 24 \mu m$.

The results of five observing seasons (one was carried out with two telescopes, M2 and S2) are reported here. Four nights out of fifteen were rejected due to observing problems like extremely bad seeing and clouds. From the remaining eleven nights of observation, one was taken with the M1, two with the M2, five with the M3, one with the S1, and two with the S2. On these data we detected  year-to-year, month-to-month, day-to-day, and intra-night variability. Only non-variable stars in the same frame were taken into consideration.

 \begin{table}
   \centering
      \caption[]{Summary of observations. See the text for a full description of the table. The magnitudes listed for December 2001 are a mean values.}
         \label{KapSou}
     $$
         \begin{array}{c c c c c c c c}
           \hline
           \hline
 Date       & Tels. & m_B & \delta m_B & m_V & \delta m_V & m_R & \delta m_R \\
\hline
 Dec/11/1998 & S1   & 16.91 & 0.02 & 16.50 & 0.01 & 15.97 & 0.02 \\

 Mar/03/2000 & M2   & 16.60 & 0.04 & 16.09 & 0.03  & 15.56 & 0.03 \\

 Mar/04/2000 & S2   & 16.61 & 0.01 & 16.10 & 0.01 & 15.60 & 0.01 \\

 Mar/05/2000 & S2   & 16.29 & 0.02 & 15.78 & 0.01 & 15.25 & 0.03 \\

 Nov/26/2000 & M3   & 17.20 & 0.02 & 16.70 & 0.01 & 16.17 & 0.01 \\

 Nov/28/2000 & M3   & 17.29 & 0.02 & 16.83 & 0.02 & 16.32 & 0.02 \\

 Nov/29/2000 & M3   & 17.30 & 0.02 & 16.80 & 0.02 & 16.26 & 0.02 \\

 Dec/03/2000 & M3   & 17.19 & 0.02 & 16.63 & 0.03 & 16.08 & 0.02 \\

 Dec/04/2000 & M3   & 16.98 & 0.02 & 16.48 & 0.02 & 15.89 & 0.02 \\

 Dec/19/2001 & M1   & 15.07 & 0.01 & 14.49 & 0.01 & 13.94 & 0.01 \\

 Apr/26/2003 & M2   & 16.90 & 0.03 & 16.36 & 0.03 & 15.74 & 0.03 \\
\hline
         \end{array}
    $$
   \end{table}

The observations were done with the strategy explained by de Diego et al. (\cite{de diego}),   taking five images with each filter in BVR sequence with approximately one hour between each observation set. The observations were reduced with the IRAF-APPHOT package. The best aperture was determined using the growth curves technique, which resulted in a pseudoaperture $\sim$3 FWHM. Standard stars given by Landolt (\cite{landolt}) were used to obtain our magnitudes. Uncertainties were determined from magnitude dispersion within each image sequence, which does not differ significantly from dispersion between the groups of images for the reference stars. When the data were transformed to fluxes, corrections to the rest frame of the object and dereddening were carried out.

We detected three timescales of variability: one year ($\sim$2.3 mag in BVR, between November 2000 and December 2001), one day ($\sim$0.1 in BVR, within the November 2000 season), and intra-night ($\sim$0.3 mag in BVR, in December 2001). These variations are studied in the next section.

\section{Analysis of variability}

\subsection{ Variability}

\object{PKS 0736+017} was observed within a larger program to study AGN microvariability. A few observations were also made to monitor long-term variability. Thus, the data permit us to obtain relevant information about short- and long-term variability. However, the sparse long-term sampling limits the possibilities for determining parameters such as the precise moment when an outburst started and finished, or the date on which the maximum occurred.

Figures 1 and 2 show the BVR light curves, which show flux variations with timescales of years and minutes, respectively. The long-term variations have amplitudes up to a few magnitudes, while the amplitudes for the short-term variability range from a few hundredths to several tenths of a magnitude. In spite of the poor time sampling, the continuous activity is obvious. Fig. 1 also shows light curves for three of our control field stars. Some stars in the \object{PKS 0736+017} field show variability of much smaller amplitude (less than 0.03 mags), but only those that remain constant during the observations were selected. In addition, star 2 from Smith et al. (\cite{smithp}) for the field of PKS~0736+017 was observed. Our magnitude estimations for this star (B=15.92, V=15.40, and R=15.04) differ slightly from those given by Smith et al. (B=15.85, V=15.37, and R=14.97).

   \begin{figure}
   \centering
   \includegraphics[trim=15.3 14 15.3 16,width=8.8cm,clip]{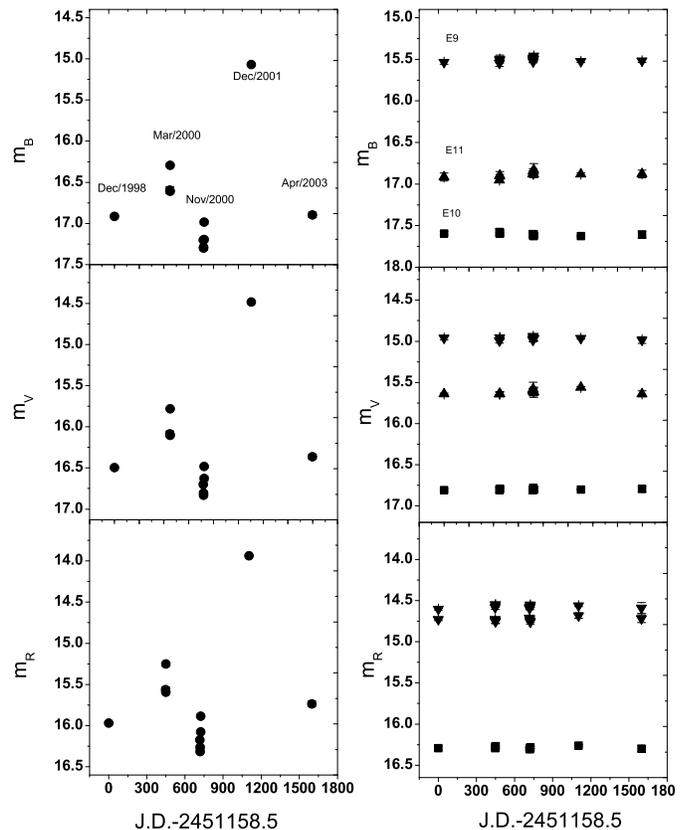}
         \label{FigCurLig}
      \caption{The left panel shows the light curves in B,V, and R bands for \object{PKS 0736+017}, as listed in Table 1. The right panel shows the reference stars included in our field: down triangles for star 9, squares for star 10, and up triangles for star 11. The December 2001 points are a mean value. J.D. = 2451158.5 corresponds to December 11, 1998.
 }
   \end{figure}

An important event was detected on December 19, 2001, when the flux level reached a maximum of B$\simeq$14.90, V$\simeq$14.34, R{\bf $\gse$}13.79 (in Tables 1 and 2, and Fig. 1 we only show mean values of the magnitudes for this night). Because of the sparse sampling, the real maximum is not well timed, but it should have happened around this epoch (see Fig. 1). Added to this outburst, \object{PKS 0736+017} showed a microvariability event during the 4.5 hours that it was monitored (Fig. 2c). It decreased its brightness at a rate of $\sim$0.066 magnitudes/hour. Due to instrumental problems we lost the first R point.

Previous photometric observations showed that this blazar varied between 16 and 17 mag in V (e.g. Garcia et al. \cite{garcia}; Bondi et al. \cite{bondi}; Wright et al. \cite{wright98}; Katajainen et al. \cite{katajainen}). Throughout our observations the brightness was generally between these historical values. In December 2001, we detected the brightest magnitude (see Table 1).

In general, the amplitudes of the variations are much larger than the uncertainties, and statistical techniques are unnecessary (the smallest detected flux variations, between November 26 and 28, 2000, are 0.09$\pm$0.03 magnitudes in B, 0.13$\pm$0.02 magnitudes in V, and 0.15$\pm$0.02 magnitudes in R).

We presume that this activity began at least in early 1998, since 
Katajainen et al. (\cite{katajainen}) reported V magnitudes in the range from 16.84 to 16.25 between 1995 and 1997. Later, Ghosh et al. (\cite{ghosh}) observed an R magnitude of 16.24 (within the typical values) on December 15, 1997, and on March 25, 1998 they detected that the brightness had increased to 15.47, and finally diminished to 15.94 in April 1998 in the same band. On December 11, 1998 we detected that the brightness had dimmed to V=16.50, and R=15.97.

The object was monitored to study its long-term variability in December 1998 and April 2003. Its brightness was similar in both epochs (V $\sim$ 16.4), although the observed colour was different, bluer in December 1998 than in April 2003 (see section 3.2 and Table 2).

 \begin{figure}
   \centering
   \includegraphics[trim= 14 13.5 13 13.5,width=8.8cm,clip]{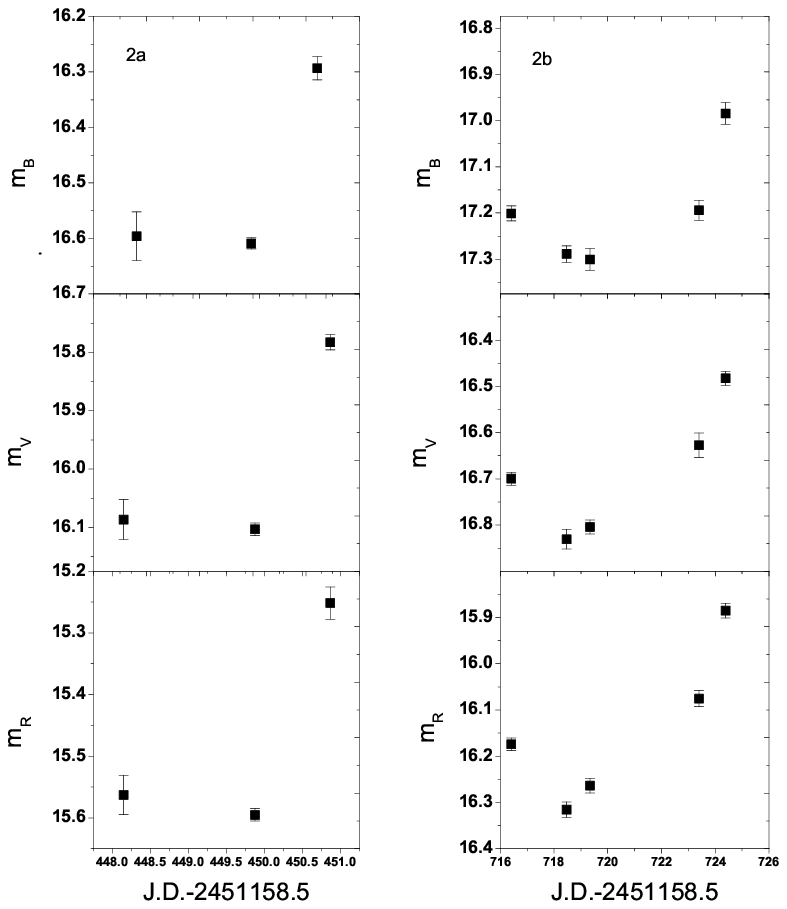}
      \caption{({\bf a-c}).{\bf a)}Light curves for the March 2000 season,  and {\bf b)} for November 2000. Magnitudes for December 2001 shown in {\bf c} are instrumental.}
         \label{FigCurLig2}
   \end{figure}

  \begin{figure}
   \centering
   \includegraphics[trim=7 7.5 0 10,width=4.9cm,clip]{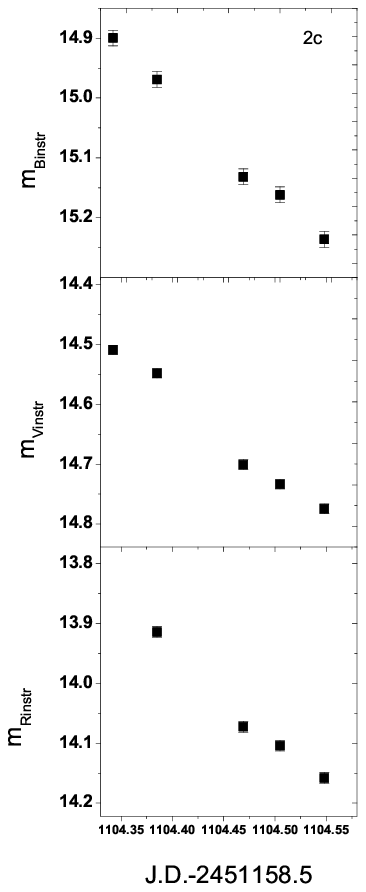}
         \label{FigCurLig2c}
   \end{figure}

 \begin{table}
 \centering
    \caption{Comparison of the spectral slope relative to the R magnitude.} 
     $$
         \begin{array}{c c c c}
         \hline
         \hline
 Date             & m_R & \alpha & \delta \alpha \\
\hline
 Dec/11/1998   & 15.97 & 1.24  & 0.05  \\

 Mar/03/2000   & 15.56 & 1.40  & 0.07  \\

 Mar/04/2000   & 15.60 & 1.38  & 0.02  \\

 Mar/05/2000   & 15.25 & 1.45  & 0.04  \\

 Nov/26/2000   & 16.17 & 1.40  & 0.04  \\

 Nov/28/2000   & 16.32 & 1.30  & 0.04  \\

 Nov/29/2000   & 16.26 & 1.41  & 0.05  \\

 Dec/03/2000   & 16.08 & 1.58  & 0.07  \\

 Dec/04/2000   & 15.89 & 1.55  & 0.05  \\

 Dec/19/2001   & 13.94 & 1.72  & 0.02  \\

 Apr/26/2003   & 15.74 & 1.67  &  0.02  \\
\hline
         \end{array}
    $$
   \end{table}

\subsection{ Variability behaviour}

The most remarkable behaviour through all our observations is that \object{PKS 0736+017} became redder when it was brighter (see Figs. 3, 4 and 5). In other words, the spectrum became steeper when the object was bright and flatter when it faded. This behaviour is opposite to the general reports for this class of objects (e.g. Kedziora-Chudczer et al. \cite{kedziora}; Gear et al. \cite{gear}; Racine 1970; Gear, Robson \& Brown \cite{gear86}; Massaro et al. \cite{massaro}; Maesano et al. \cite{maesano}; Papadakis et al. \cite{papadakis}), and common models (e.g. Li and Kusunose \cite{li}; Chiang and B\"ottcher \cite{chiang}; Wang and Kusunose \cite{wang}; Dermer \cite{dermer}; Spada et al. \cite{spada}; Sikora et al. \cite{sikora}; Qian et al. \cite{qian}; Wiita \cite{wiita}; Dermer \& Schlickeiser \cite{dermer02}; Nesci et al. \cite{nesci}; Marscher et al. \cite{marscher80}; Tr\`evese \& Vagnetti \cite{trevese}) which state that blazars become bluer when they brighten. This peculiar behaviour was also noted by Clements et al. (\cite{clements}) in early 2002, just a month after the flare of December 2001. These authors reported V=14.322 and R=13.717 (V-R $\simeq$0.605) at the maximum. They concluded that the phenomenon ``appears to be more related to the nature of the variation than to the host galaxy or the brightness level''.

   \begin{figure}
   \centering
          \includegraphics[trim= 7 11 7 9,width=8cm,clip]{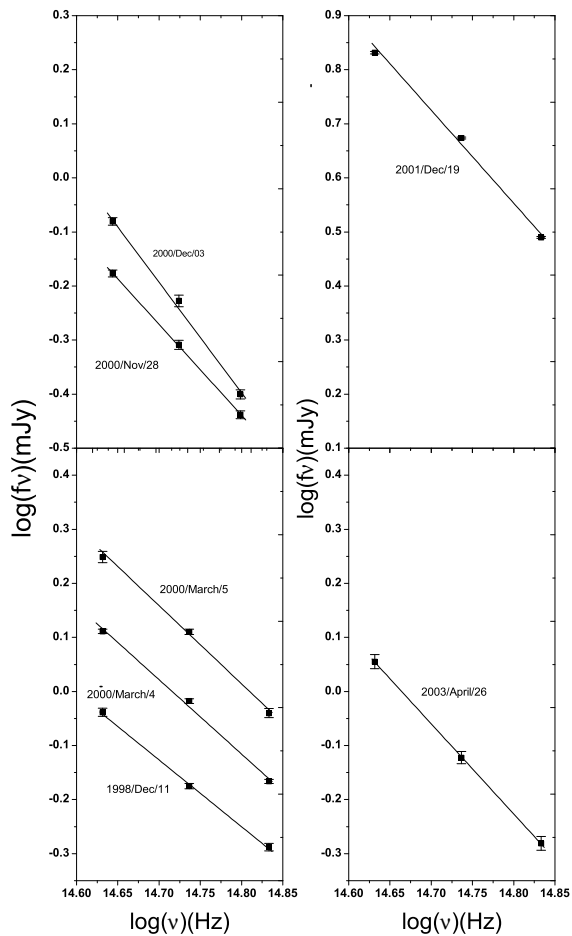}
      \caption{The shape of the optical broad band of \object{PKS 0736+017} can be fitted by a power law.}
         \label{FigSpeSlo}
   \end{figure}

Our data can be described well with a power law, i.e., f$_{\nu} \propto \nu ^{-\alpha}$, where ${-\alpha}$ is the spectral slope  in a log flux vs. log frequency diagram (see Fig. 3), which permits us to quantify the reddening accurately. Better fits can be made if more spectral components are included in the spectrum model, such as the contribution of the galaxy (e.g. Wright et al. \cite{wright98}) or a blackbody (e.g. Malkan and Moore \cite{malkan}), but with spectral resolution of only three points an extra component does not make sense.

The values of the spectral slope and their uncertainties for each epoch are given in columns (3) and (4) in Table 2. In this table the date and the R magnitude are listed in columns (1) and (2), respectively. It is evident that commonly the brightest points have steeper spectra. This tendency was noticed in November 2000, when we detected a slope steepening from $\alpha=$1.30 to 1.55 while the flux increased (Fig. 4b). Furthermore, the flux and the spectral slope are well correlated, with a correlation coefficient of 89~\%. In March 2000 we have two estimates of $\alpha$ (1.38 and 1.45) and one of them with a large uncertainty. However, Fig. 4a shows that the tendency is still the same as in Fig. 4b. At this time, the brightness increased while the slope changed slightly. Note that the uncertainties for March 2000 do not overlap.

A similar tendency to redden with increased brightness was observed in the long-term variability (see Fig. 5), but a marginal correlation between spectral index and flux level was found. Some data differ from the common behaviour in our observations. In particular, in December 2001, while the flux decreased by 0.3 mags the spectral slope remained constant ($\alpha$=1.72). On the other hand, April 2003 data has a low flux level and high spectral slope value. Nevertheless, we can find an explanation in the online data of Clements et al. (\cite{clements}). On an $\alpha$-log flux plane, these data are distributed in two \emph{varying modes}: one producing larger flux variations with shorter spectral changes (tendency represented by dashed line in Fig. 5), and other one with the inverse effect (tendency represented by dotted line in Fig. 5). Then, December 2001 could be understood if it is varying as the last variable mode, while April 2003 as the former one.

   \begin{figure}
   \centering
   \includegraphics[trim= 10 15 10 9,width=8.8cm,clip]{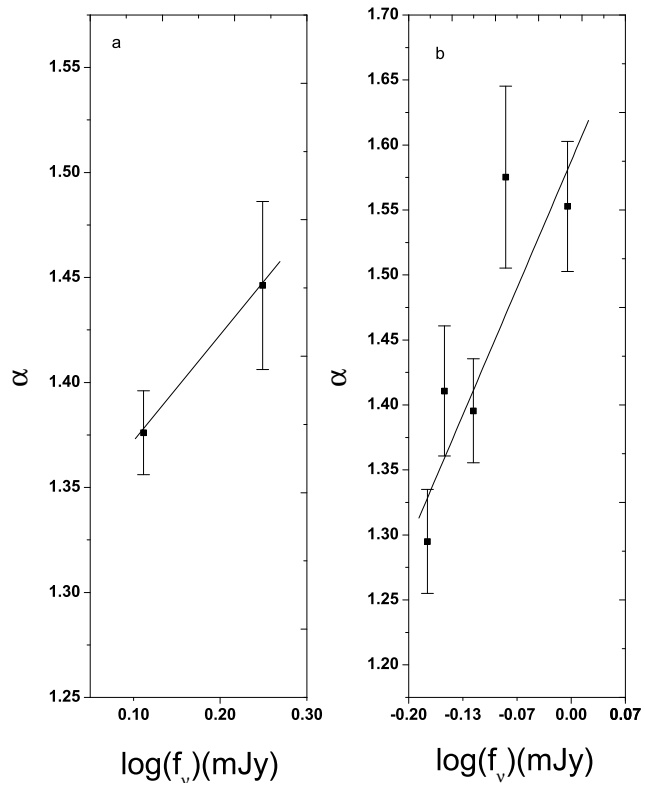}
      \caption{{\bf a)} Data for March 2000 shows a tendency in the sense that \object{PKS 0736+017} is redder when it is brighter. {\bf b)} This tendency was confirmed with data obtained in November 2000, which shows a clear correlation between colour and brightness.}
         \label{FigCorSlo}
   \end{figure}

In December 2001 both flux level and behaviour were similar to those reported by Clements et al. (\cite{clements}) for the January 2002 flare. In Fig.5, Clements et al.'s average maximum from January 23 is plotted (roughly computed with only R and V bands).

\section{Discussion}

We stated that the spectrum can be described well by a power law, although other spectral components could be included. For example, Malkan and Moore (\cite{malkan}) observed \object{PKS 0736+017} at ultraviolet, optical and infrared bands. In the optical and ultraviolet bands they fitted a power law with spectral index of 1.0 plus a blackbody with a temperature of 26,000 K, but the latter should only contribute a fourth of the total brightness. Additionally, these authors suggested that \object{PKS 0736+017} has combined emission from a ``normal'' quasar (with a mixture of both power law emission with a spectral slope of 1.1, and thermal emission from accretion disk) and a blazar emission (which is represented by a power law with a slope between 1.1 and 2.0). They further proposed that the blazar component should be increasingly important at longer wavelengths, and when the continuum is brighter.

Brown et al. (\cite{browna}) also proposed that the emission of OVV quasars is a mixture of a ``quiescent'', a violently variable and a thermal emission component. The first component should be responsible for the centimetre emission, the second one dominates the millimetre to ultraviolet region, and the third is responsible for some of the optical and ultraviolet emission. With this picture in mind, the violently variable component should cause the flares and outbursts in these objects. Brown et al. (\cite{browna},\cite{brownb}) also mentioned that, at high brightness state, the spectral shape of OVV quasars showed similar properties to the emission of BL Lac objects.

These simple models could give a qualitative explanation only for our low flux level data. This can be noticed considering that, the composite spectrum is flatter than the non-thermal component, because the thermal contribution is larger in the blue part. Then, when the object is brightening, the non-thermal component has an even more dominant contribution to the total flux, and the composite spectrum steepens.

 \begin{figure}
   \centering
   \includegraphics[trim= 27 27 10 16,width=8.8cm,clip]{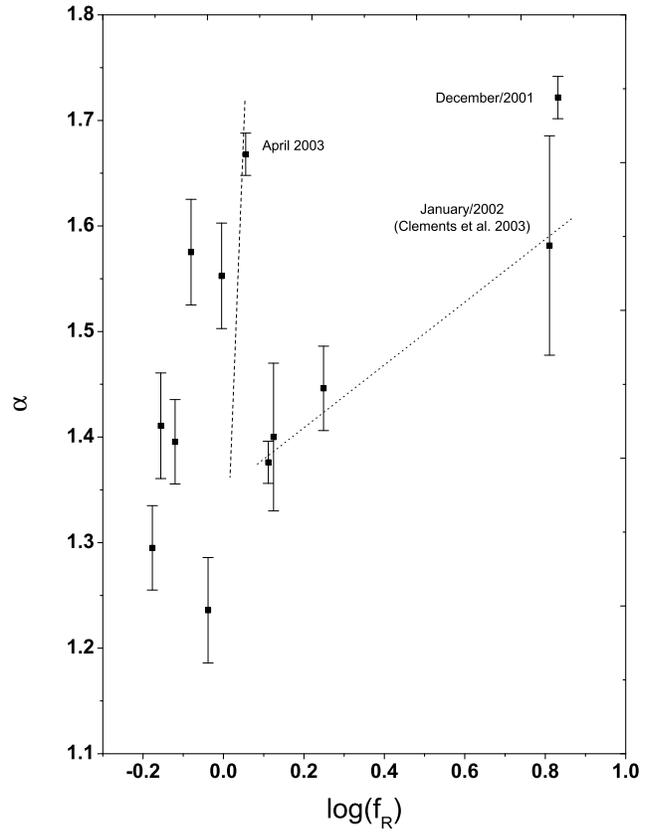}
      \caption{Relation between the spectral index $\alpha$ and the brightness in R. We have included our own data and the average maximum reported by Clements et al. (2003) for January 2002. Our analysis of the data reported by Clements et al., in their article online version, suggests two varying modes. This tendency is represented by the dashed and dotted lines. Our data agrees with these tendency. At low flux level, small changes in flux correspond to large changes in the spectral slope, while much less pronounced spectral changes correspond to high brightness state. In both cases, the object reddens when it brightens. }
         \label{FigCorSloall}
   \end{figure}

However, high flux levels have no satisfactory explanation because our point of largest slope, in December 2001, corresponds to an outburst (at this time, the \emph{blazar component} must be dominant and the flattest spectra must be measured, as is observed in these objects). The bulk of our data differ from synchrotron loss energy behaviour (Brown et al. \cite{browna}). A thermal varying bump cannot explain our data because changes of this component lead to larger changes in the blue part of the spectrum. Additionally, two variability modes in the $\alpha$-log flux plane are observed.

We have detected a behaviour that is not explained by the usual mechanisms. The evidence that the amplitudes of variations are not systematically larger at high frequencies in this and other blazars is increasing (see for example Ghosh et al. \cite{ghosh}; Brown et al. \cite{brownb}; Malkan and Moore \cite{malkan}; Massaro et al. \cite{massaro}; Clements et al. \cite{clements}, and this work).

Brown et al. (\cite{brownb}) observed \object{PKS 0736+017} at infrared wavelengths and found no correlation between spectral slope and flux level; however, if they disregard one point at the minimum, they find that the spectrum steepens when it is brighter, as they also found in the case of the blazar \object{3C 279}. They did not find a significant correlation between the flux level and the slope values for other three blazars (\object{1308+326}, \object{3C 345}, and \object{3C 446}). They point out that this absence of correlation is not consistent with the behaviour expected from emission of a single synchrotron component, therefore other components must be involved.

Earlier, Miller (\cite{miller}) found that the blazar \object{3C 446} became redder as the source brightened. Brown et al. (\cite{brown}) reported that the spectrum of \object{3C 446} steepened with brightening in the K-band in July 1980 and September 1983. They explained their observations with the argument that the maximum of brightness could have occurred before their observations and that the spectrum steepened rapidly. This explanation cannot be applied to our data because we observed \object{PKS 0736+017} when it was brightening (November 2000).

Ghosh et al. (\cite{ghosh}) found a reddening in their optical observations of the BL Lac object \object{PKS~0735+178}. Additionally, they observed that in the blazar \object{PKS~0235+164}, the spectral slope remained almost constant when its brightness increased. They concluded that these characteristics cannot be described by  energy losses in a pure synchrotron mechanism scenario (one of the most used flare models).

\section{ Conclusions}

The OVV quasar \object{PKS 0736+017} was observed between December 1998 and April 2003. Variability events were detected throughout the observations at different timescales and amplitudes. Some flares were detected in March 2000 and November-December 2000, with flux change of $\sim$0.32 and $\sim$0.4 mags, respectively.

The most important features that we report are:

$\bullet$ An outburst was detected on December 19, 2001.  At this time, the bright level reached a maximum of 14.90 in the B band, 14.34 in the V band, and 13.79 in the R band. A brightness decrease of $\sim$0.3 mags occurred in 4.5 hours, while the spectral slope was unchanged.

$\bullet$ A tendency to redden with increased brightness was detected throughout our observations. Moreover, a good correlation between flux level and spectral slope was detected in November 2000, while a clear tendency was found in March 2000. Clements et al. (2003) present data strongly suggesting the existence of two varying modes. Our data agree with this assumption.

Clements et al. (\cite{clements}) also detected the same tendency between brightness and spectral slope for \object{PKS 0736+017}, but other authors did not (e.g. Brown et al. \cite{brownb}; Malkan \& Moore \cite{malkan}). There is no doubt that more observations of blazars that show reddening as they brighten are needed. Multiwavelength observations with good time resolution on targets like \object{3C 446}, \object{PKS 0735+178}, and \object{PKS~0736+017} will give us important information about the physical conditions of these objects. Models should be improved to account for these observational results.

\begin{acknowledgements}
This work was supported by Grant No. 121553 from CONACyT, and PAPIIT projects IN115599, and IN113002. The Jacobus Kapteyn Telescope is operated on the island of La Palma by the Isaac Newton Group in the Spanish Observatorio del Roque de los Muchachos of the Instituto de Astrof\'\i sica de Canarias. We acknowledge the use of IRAF, distributed by NOAO. We also acknowledge to Garc\'\i a Ruiz, G., Melgoza, G., Monrroy, S., and Montalvo, F. for their support during the observations. 

\end{acknowledgements}


\begin{thebibliography}{}
\bibitem[1980]{angel}
 Angel, J.R.P. \& Stockman, H.S. 1980, ARA\&A, 18, 321
\bibitem[1985]{antonucci} 
 Antonucci, R.R.J. \& Ulvestad, J.S. 1985, ApJ, 294, 158
\bibitem[1978]{blanford} 
 Blanford, R.D. \& Rees, M.J. 1978, in Wolfe A.M. ed. Pittsburgh Conference on BL Lac Objects. University of Pittsburgh, p.328
\bibitem[1996]{bondi}
 Bondi, M., Padrielli, L., Fanti, R. et al. 1996, A\&A, 308, 415
\bibitem[1986]{brown}
 Brown, L.M.J., Robson, E.I., Gear, W.K. et al. 1986, MNRAS, 219, 671
\bibitem[1989a]{browna}
 Brown, L.M.L., Robson, E.I, Gear, W.K.  et al. 1989a, ApJ, 340, 129
\bibitem[1989b]{brownb}
 Brown, L.M., Robson E.I., Gear, W.K. \& Smith, M.G. 1989b, ApJ, 340, 150
\bibitem[2002]{chiang}
 Chiang, J. \& B\"ottcher, M. 2002, ApJ, 564, 92
\bibitem[1996]{cristiani}
 Cristiani, S., Trentini, S., La Franca, S. et al. 1996, A\&A, 306, 395
\bibitem[2003]{clements}
 Clements, S.D., Jenks, A. \& Torres, Y. 2003, AJ, 126, 37
\bibitem[1998]{de diego}
 de Diego, J.A., Dultzin-Hacyan, D., Ram\'\i rez, A. \& Ben\'\i tez, E. 1998, ApJ, 500, 69
\bibitem[1998]{dermer}
 Dermer, C.D. 1998, ApJ, 501, L157
\bibitem[2002]{dermer02}
 Dermer, C.D. \& Schlickeiser, R. 2002, ApJ, 575, 667
\bibitem[1996]{diclemen}
 Di Clemente, A., Giallongo, E., Natali, G., Trevese, D. \& Vagnetti, F. 1996, ApJ, 463, 466
\bibitem[1993]{dunlop}
 Dunlop, J.S., Taylor, G.L., Hughes, D.H. \& Robson, E.I. 1993, MNRAS, 264, 455
\bibitem[1994]{falomo}
 Falomo, R., Scarpa, R. \& Bersanelli, M. 1994, ApJS, 93, 125
\bibitem[2000]{falomo00}
 Falomo, R. \& Ulrich, M.H. 2000, A\&A, 357, 91
\bibitem[1999]{garcia}
 Garcia, A., Sodr\'e, L., Jablonski, F.J. \& Terlevich, R.J. 1999, MNRAS, 309, 803
\bibitem[1986]{gear}
 Gear, W.K., Brown, L.M.J., Robson, E.I. et al. 1986, ApJ, 304, 295
\bibitem[1986]{gear86}
 Gear, W.K., Robson, E.I. \& Brown, L.M.J. 1986, Nature, 324, 546
\bibitem[2002]{ghisellini02}
 Ghisellini, G., Celotti, A., Costamante, L. 2002, A\&A, 386, 833
\bibitem[1997]{ghisellini}
 Ghisellini, G., Villata, M., Raiteri, C.M. et al. 1997, A\&A, 327, 61
\bibitem[2000]{ghosh}
 Ghosh, K.K., Ramsey, B.D., Sadun, A.C. \& Soundararaja-perumal, S. 2000, ApJS, 127, 11
\bibitem[1994]{hook}
 Hook, I.M., McMahon. R.G., Boyle, B.J. \& Irwin, M.J. 1994, MNRAS, 268, 305
\bibitem[2000]{hughes}
 Hughes, D., Kukula, M.J., Dunlop, J.S. \& Boroson, T. 2000, MNRAS, 316, 204
\bibitem[1987]{hutchings}
 Hutchings, J.B. 1987, ApJ, 320, 122
\bibitem[1988]{hutchings88}
 Hutchings, J.B., Johnson, I. \& Pyke, R. 1988, ApJS, 66, 361
\bibitem[1981]{jones}
 Jones, W.T., Rudnick, L., Owen, F.N. et al.  1981, ApJ, 243, 97
\bibitem[2000]{katajainen}
 Katajainen, S., Takalo, L.O., Sillamp\"a\"a, A. et al. 2000, A\&A, 143, 357
\bibitem[2001]{kedziora}
 Kedziora-Chudczer, L.L., Jauncey, D.L., Wieringa, M.H., Tzioumis, A.K. \& Reynolds, J.E. 2001, MNRAS, 325, 1411
\bibitem[1998]{kotil}
 Kotilainen, J., Falomo, R. \& Scarpa, R. 1998, A\&A, 332, 503
\bibitem[1992]{landolt}
Landolt, A.U. 1992, AJ, 104, 340
\bibitem[2000]{li}
 Li, H. \& Kusunose, M. 2000, ApJ, 536, 729
\bibitem[1997]{maesano}
 Maesano, M., Messaro, E. \& Nesci, R. 1997, IAU, Circ. 6700
\bibitem[1997]{maesano97}
 Maesano, M., Montagni, F., Massaro, E. \& Nesci, R. 1997, A\&A, 122, 267
\bibitem[1986]{malkan}
 Malkan, M. \& Moore, R. 1986, ApJ, 300, 216
\bibitem[1993]{mangalam}
 Mangalam, A.V. \& Wiita, P.J. 1993, ApJ, 406, 420
\bibitem[1994]{maraschi}
 Maraschi, L., Grandi, P., Urry, C.M. et al. 1994, ApJ, 435, L91
\bibitem[1989]{maraschi89}
 Maraschi, L., Celotti, A. \& Treves, A. 1989, in Proc. 23d, ESLAB Symp. on Two Topics in X-ray Astronomy ed. J.J. Hunt \& B. Battrick (vol. 2; Paris: ESA), 825
\bibitem[1996]{marscher}
 Marscher, A.P. 1996, in ASP Conf. Ser. 110, Blazar continuum Variability, ed. H.R. Miller, J.R. Webb \& J.C. Noble (San Francisco: ASP), 248
\bibitem[1992]{marscher92}
 Marscher, A.P., Gear, W.K. \& Travis, J.P. 1992, in Variability of Blazars, ed. E. Valtaoja \& M. Valtonen (Cambridge Univ. Press), 85
\bibitem[1980]{marscher80}
 Marscher, A.P. 1980, ApJ, 235, 386 
\bibitem[1998]{massaro}
 Massaro, E., Nesci, R., Maesano, M., Montagni, F. \& D'Allessio, F. 1998, MNRAS, 299, 47
\bibitem[1976]{mcadam}
 McAdam, W.B. 1976, PASA, 3, 86
\bibitem[1975]{mcgimsey}
 McGimsey, B.Q., Smith, A.G., Scott, R.L., et al. 1975, AJ, 80, 895
\bibitem[1981]{miller}
Miller, H.R. 1981, ApJ, 244, 426
\bibitem[1997]{nair}
 Nair, A.D. 1997, MNRAS, 287, 641
\bibitem[1998]{nesci}
 Nesci, R., Maesano, M., Massaro, E. et al. 1998, A\&A, 332, L1
\bibitem[1996]{netzer}
 Netzer, H., Heller, A., Loinger, F. et al. 1996, MNRAS, 279, 429
\bibitem[1992]{padovani}
 Padovani, P. \& Urry, C.M. 1992, ApJ, 387, 449
\bibitem[1987]{padrielli}
 Padrielli, L., Aller, M.F., Aller, H.D. et al. 1987, A\&AS, 67, 63
\bibitem[2003]{papadakis}
 Papadakis, I.E., Boumis, P., Samaritakis, V. \& Papamastora-kis, J. 2003, A\&A, 397, 565
\bibitem[1980]{pica}
 Pica, A.J., Pollock, J.T., Smith, A.G. et al. 1980, AJ, 85, 1442
\bibitem[1988]{pica88}
 Pica, A.J., Smith, A.G., Webb, J.R. et al. 1988, AJ, 96, 1215
\bibitem[1991]{qian}
 Qian, S.J., Quirrembach, A., Witzel, A. et al. 1991, A\&A, 241, 15
\bibitem[1970]{racine}
 Racine, R. 1970, ApJ, 159, L99
\bibitem[2003]{ramirez}
 Ram\'\i rez, A., de Diego, J.A., Dultzin-Hacyan, D. \& Ben\'\i tez, E., 2003, in ``Active Galaxy Nuclei: From central engine to Host Galaxy'', ASP Conference Series, Vol. 290, 2003, eds. S.Collin, F. Combes, and I. Shlosman, p. 123
\bibitem[1980]{schaefer}
 Schaefer, B.E. 1980, PASP, 92, 255
\bibitem[2001]{sikora}
 Sikora, M., Blazejowski, M., Begelman, M.C. \& Moderski, R. 2001, ApJ, 554, 1
\bibitem[1985]{sitko}
 Sitko, M.J., Schmidt, G.D. \& Stein, W.A. 1985, ApJS, 59, 323
\bibitem[1986]{smith}
 Smith, E.P., Heckman, T.M., Bothun, G.D., Romanishin, W. \& Balick, B. 1986, ApJ, 306, 64
\bibitem[1985]{smithp}
Smith, P.S., Balonek, T.J., Heckert, P.A. et al. 1985, AJ, 90, 1184
\bibitem[2001]{spada}
 Spada, M., Ghisellini, G., Lazzati, D. \& Celotti, A. 2001, MNRAS, 325, 1559
\bibitem[2002]{trevese}
 Tr\`evese, D. \& Vagnetti F. 2002, ApJ, 564, 624
\bibitem[2003]{vagnetti}
 Vagnetti, F., Trevese, D., \& Nesci, R. 2003, ApJ, 590, 123
\bibitem[1985]{wall}
 Wall, J.V. \& Peacock, J.A. 1985, MNRAS, 216, 173
\bibitem[2002]{wang}
 Wang, J.-M. \& Kusunose, M. 2002, ApJS, 138, 249
\bibitem[1996]{wiita}
 Wiita, P.J. 1996, in ASP Conf. Ser. 110, Blazar continuum Variability, ed. H.R. Miller, J.R. Webb \& J.C. Noble (San Francisco: ASP), 42
\bibitem[1984]{wright}
 Wright, A.E. 1984, PASA, 5, 510
\bibitem[1998]{wright98}
 Wright, S.C., McHardy, I.M. \& Abraham, R.G. 1998, MNRAS, 295, 799
\bibitem[1981]{wyckoff}
 Wyckoff, S., Gehren, T. \& Wehinger, P.A. 1981, ApJ, 247, 750
\end{thebibliography}
\end{document}